# Fast Bayesian Intensity Estimation for the Permanental Process

**Christian J. Walder** [1 2] **Adrian N. Bishop** [1 2 3]

## Abstract

The Cox process is a stochastic process which generalises the Poisson process by letting the underlying intensity function itself be a stochastic process. In this paper we present a fast Bayesian inference scheme for the permanental process, a Cox process under which the square root of the intensity is a Gaussian process. In particular we exploit connections with reproducing kernel Hilbert spaces, to derive efficient approximate Bayesian inference algorithms based on the Laplace approximation to the predictive distribution and marginal likelihood. We obtain a simple algorithm which we apply to toy and real-world problems, obtaining orders of magnitude speed improvements over previous work.

## 1. Introduction

The Poisson process is an important model for point data in which samples of the process are locally finite subsets of some domain such as time or space. The process is parametrised by an *intensity function*, the integral of which gives the expected number of points in the domain of integration — for a gentle introduction we recommend (Baddeley, 2007). In the typical case of unknown intensity function we may place a non-parametric prior over it via *e.g.* the Gaussian Process (GP) and perform Bayesian inference.

Inference under such models is challenging due to both the GP prior and the non factorial nature of the Poisson process likelihood (1), which includes an integral of the intensity function. One may resort to discretising the domain (Rathbun & Cressie, 1994; Møller et al., 1998; Rue et al., 2009) or performing Monte Carlo approximations (Adams et al., 2009; Diggle et al., 2013). Fast Laplace approximates were studied in (Cunningham et al., 2008; Illian et al., 2012; Flaxman et al., 2015) and variational methods were applied in (Lloyd et al., 2015; Kom Samo & Roberts, 2015).

To satisfy non-negativity of the intensity function one transforms the GP prior. The *log-Gaussian Cox Process*, with GP distributed log intensity, has been the subject of much study; see e.g. (Rathbun & Cressie, 1994; Møller et al., 1998; Illian et al., 2012; Diggle et al., 2013), Alternative formulations for introducing a GP prior exist, *e.g.* (Adams et al., 2009). More recent research has highlighted the analytical and computational advantages (Lloyd et al., 2015; 2016; Flaxman et al., 2017; Møller et al., 1998) of the *permanental process*, which has GP distributed square root intensity (Shirai & Takahashi, 2003; McCullagh & Møller, 2006) — we discuss the relationship between these methods and the present work in more detail in subsection 2.2.

In section 2 we introduce the Poisson and permanental processes, and place our work in the context of existing literature. Section 3 reviews Flaxman et al. (2017), slightly recasting it as regularised maximum likelihood for the permanental process. Our Bayesian scheme is then derived in section 4. In section 5 we discuss the choice of covariance function for the GP prior, before presenting some numerical experiments in section 6 and concluding in section 7.

## 2. The Model

### 2.1. The Poisson Process

We view the inhomogeneous Poisson process on $\Omega$ as a distribution over locally finite subsets of $\Omega$. The number $N(\mathcal{X})$ of elements in some $\mathcal{X} \subseteq \Omega$ is assumed to be distributed as $\text{Poisson}(\Lambda(\mathcal{X}, \mu))$, where $\Lambda(\mathcal{S}, \mu) := \int_{\boldsymbol{x} \in \mathcal{S}} \lambda(\boldsymbol{x}) d\mu(\boldsymbol{x})$ gives the mean of the Poisson distribution. It turns out that this implies the likelihood function

$$p\left(\{\boldsymbol{x}_i\}_{i=1}^m | \lambda, \Omega\right) = \prod_{i=1}^m \lambda(\boldsymbol{x}_i) \exp\left(-\Lambda(\Omega)\right). \quad (1)$$

### 2.2. Latent Gaussian Process Intensities

To model unknown $\lambda(\boldsymbol{x})$, we employ a non-parametric prior over functions, namely the Gaussian process (GP). To ensure that $\lambda$ is non-negative valued we include a deterministic "link" function $g : \mathbb{R} \to \mathbb{R}^+$ so that we have the prior over $\lambda$ defined by $\lambda = g \circ f$ and $f \sim \text{GP}(k)$, where $k$ is the covariance function for $f$. The most com-mon choice for $g$ is the exponential function $\exp(\cdot)$, leading to the *log-Gaussian Cox process* (LGCP) (Møller et al., 1998). Recently Adams et al. (2009) employed the transformation $g(z) = \lambda^*(1+\exp(-z))^{-1}$ , which permits efficient sampling via *thinning* (Lewis & Shedler, 1979) due to the bound $0 \leq \lambda(\boldsymbol{x}) \leq \lambda^*$.

[1]Data61, CSIRO, Australia [2]The Australian National University [3]University of Technology Sydney. Correspondence to: Christian <christian.walder@anu.edu.au>.

#### 2.2.1. Permanental Processes: Squared Link Function

In this paper we focus on the choice $g(z) = \frac{1}{2}z^2$, known as the permanental process (Shirai & Takahashi, 2003; McCullagh & Møller, 2006). Two recent papers have demonstrated the analytical and computational advantages of this link function.

1. Flaxman et al. (2017) derived a non-probabilistic regularisation based algorithm which we review in section 3, and which exploited properties of reproducing kernel Hilbert spaces. The present work generalises their result, providing probabilistic predictions and Bayesian model selection. Our derivation is by necessity entirely different to Flaxman et al. (2017), as their representer theorem (Schölkopf et al., 2001) argument is insufficient for our probabilistic setting (see *e.g.* subsubsection 4.1.6).

2. (Lloyd et al., 2015) derived a variational approximation to a Bayesian model with the squared link function, based on an *inducing variable* scheme similar to (Titsias, 2009), and exploiting the tractability of certain required integrals. The present work has the advantage of 1) not requiring the inducing point approximation, 2) being free of non-closed form expressions such as their $\tilde{G}$ and 3) being simpler to implement and orders of magnitude faster in practice while, as we demonstrate, exhibiting comparable predictive accuracy.

## 3. Regularised Maximum Likelihood

Flaxman et al. (2017) combined (1) with the regularisation term $\|f\|^2_{\mathcal{H}(k)}$, leading to the regularised maximum likelihood estimator for $f$, namely $\hat{f} :=$

$$\underset{f}{\operatorname{argmax}} \sum_{i=1}^{m} \log \frac{1}{2} f^2(\boldsymbol{x}_i) - \frac{1}{2} \underbrace{\left( \|f\|^2_{L^2(\Omega,\mu)} + \|f\|^2_{\mathcal{H}(k)} \right)}_{:=\|f\|^2_{\mathcal{H}(k,\Omega,\mu)}}, \tag{2}$$

where we have implicitly defined the new RKHS $\mathcal{H}(k,\Omega,\mu) := \mathcal{H}(\tilde{k})$. Now, provided we can compute the associated new reproducing kernel $\tilde{k}$, then we may appeal to the representer theorem (Kimeldorf & Wahba, 1971) in order to compute the $\hat{f}$, which takes the form $\sum_{i=1}^{m} \alpha_i \tilde{k}(\boldsymbol{x}_i, \cdot)$ for some $\alpha_i$. The function $\tilde{k}$ may be expressed in terms of the Mercer expansion (Mercer, 1909)

$$k(\boldsymbol{x},\boldsymbol{y}) = \sum_{i=1}^{N} \lambda_i \phi_i(\boldsymbol{x})\phi_i(\boldsymbol{y}), \tag{3}$$

where $\phi_i$ are orthonormal in $L^2(\Omega,\mu)$. To satisfy for arbitrary $f = \sum_i w_i \phi_i$ the reproducing property (Aronszajn, 1950)

$$\left\langle k(\boldsymbol{x},\cdot), \sum_i w_i \phi(\cdot)_i \right\rangle_{\mathcal{H}(k)} := f(\boldsymbol{x}) = \sum_i w_i, \phi_i(\boldsymbol{x}) \tag{4}$$

we let $\phi_i$ be orthogonal in $\mathcal{H}(k)$, obtaining $\langle \phi_i, \phi_j \rangle = \delta_{ij}\lambda_i^{-1}$. Hence, $\|\sum_i w_i \phi_i\|^2_{\mathcal{H}(k)} = \sum_i w_i^2/\lambda_i$, and from (2) we have $\|\sum_i w_i \phi_i\|^2_{\mathcal{H}(k,\Omega,\mu)} = \sum_i w_i^2(1+\lambda_i^{-1})$, so

$$\tilde{k}(\boldsymbol{x},\boldsymbol{y}) = \sum_{i=1}^{N} \frac{1}{1+\lambda_i^{-1}} \phi_i(\boldsymbol{x})\phi_i(\boldsymbol{y}). \tag{5}$$

For approximate Bayesian inference however, we cannot simply appeal to the representer theorem.

## 4. Approximate Bayesian Inference

In subsection A.3 of the supplementary material, we review the standard Laplace approximation to the GP with non-Gaussian likelihood. This a useful set-up for what follows, but is not directly generalisable to our case due to the integral in (1). Instead, in subsection 4.1 we now take a different approach based on the Mercer expansion.

### 4.1. Laplace Approximation

It is tempting to naïvely substitute $\tilde{k}$ into subsection A.3 of the supplementary material, and to neglect the integral part of the likelihood. Indeed, this gives the correct approximate predictive distribution. The marginal likelihood does not work in this way however (due to the log determinant in (17)). We now perform a more direct analysis.

#### 4.1.1. Mercer Expansion Setup

Mercer's theorem allows us to write (3), where for non-degenerate kernels, $N = \infty$. Assume a linear model in $\Phi(\boldsymbol{x}) = (\phi_i(\boldsymbol{x}))_i$ so that[1]

$$f(\boldsymbol{x}) = \boldsymbol{w}^\top \Phi(\boldsymbol{x}), \tag{6}$$

and let $\boldsymbol{w} \sim \mathcal{N}(0,\Lambda)$ where $\Lambda = (\lambda_i)_{ii}$ is a diagonal covariance matrix. This is equivalent to $f \sim \mathrm{GP}(k)$ because

$$\operatorname{cov}(f(\boldsymbol{x}), f(\boldsymbol{z})) = \Phi(\boldsymbol{x})^\top \Lambda\, \Phi(\boldsymbol{z}) = k(\boldsymbol{x},\boldsymbol{z}).$$

[1] We use a sloppy notation where $(\boldsymbol{x})_i$ is the $i$-th element of $\boldsymbol{x}$ while $(x_i)_i$ is a vector with $i$-th element $x_i$, *etc.*

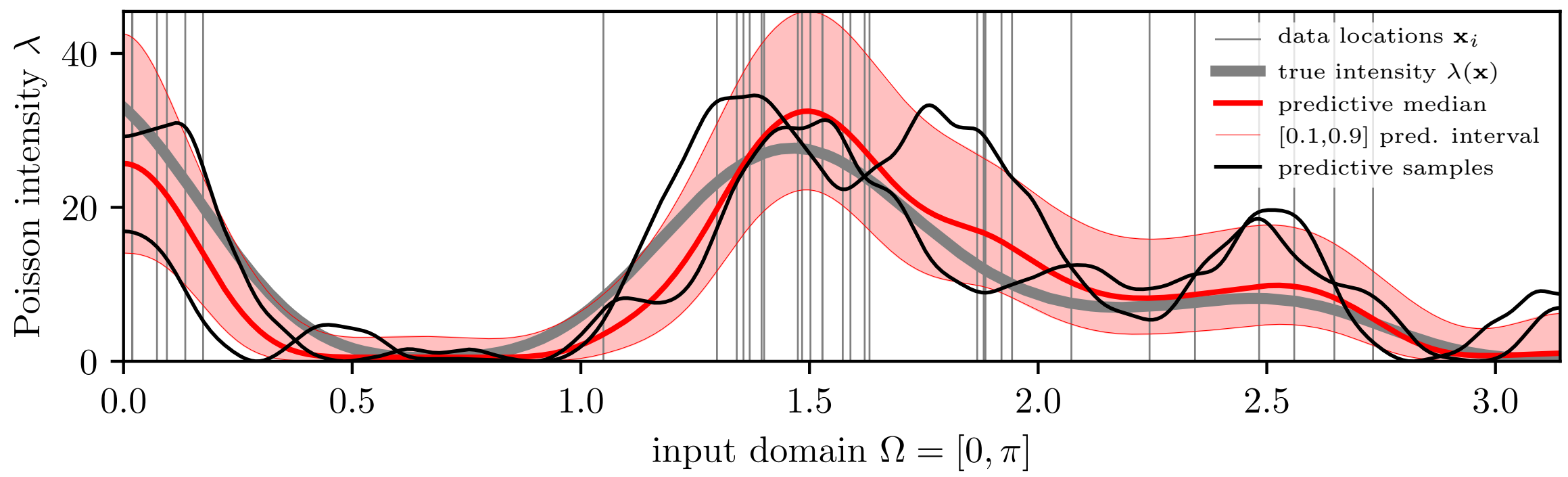

*Figure 1.* Test function $\lambda_0$ of subsection 6.2. The vertical grey lines are the input data points, sampled from the ground truth intensity depicted by the heavy grey line. We plot two samples from the approximate predictive distribution (black lines) along with the median (solid red) and $[0.1, 0.9]$ interval (filled red). The GP prior is that of subsection 5.1 with hyper-parameters $m = 2$, number of cosine frequencies $N = 256$, and the remaining $a$, $b$ chosen to maximise the marginal likelihood.

Recall that the Poisson process on $\Omega$ with intensity $\lambda(\boldsymbol{x}) = \frac{1}{2} f^2(\boldsymbol{x})$ has likelihood for $X := \{\boldsymbol{x}_i\}_{i=1}^m$

$$\log p(X|\boldsymbol{w}, \Omega, k) = \underbrace{\sum_{i=1}^m \log \frac{1}{2} f^2(\boldsymbol{x}_i)}_{:=\log h(X|\boldsymbol{w})} - \frac{1}{2} \underbrace{\int_{\boldsymbol{x}\in\Omega} f^2(\boldsymbol{x}) d\mu(\boldsymbol{x})}_{\boldsymbol{w}^\top \boldsymbol{w}}$$

The joint in $\boldsymbol{w}, X$ is

$$\log p(\boldsymbol{w}, X|\Omega, k) = \log h(X|\boldsymbol{w}) - \frac{1}{2} \boldsymbol{w}^\top (I + \Lambda^{-1}) \boldsymbol{w} - \frac{1}{2} \log |\Lambda| - \frac{N}{2} \log 2\pi.$$

### 4.1.2. Laplace Approximation

We make a Laplace approximation to the posterior, which is the normal distribution

$$\log p(\boldsymbol{w}|X, \Omega, k) \approx \log \mathcal{N}(\boldsymbol{w}|\hat{\boldsymbol{w}}, Q) \tag{7}$$
$$= -\frac{1}{2} (\boldsymbol{w} - \hat{\boldsymbol{w}})^\top Q^{-1} (\boldsymbol{w} - \hat{\boldsymbol{w}}) - \frac{1}{2} \log |Q| - \frac{N}{2} \log 2\pi$$
$$:= \log q(\boldsymbol{w}|X, \Omega, k),$$

where $\hat{\boldsymbol{w}}$ is chosen as the mode of the true posterior, and $Q$ is the inverse Hessian of the true posterior, evaluated at $\hat{\boldsymbol{w}}$.

### 4.1.3. Predictive Mean

The mode $\hat{\boldsymbol{w}}$ is

$$\hat{\boldsymbol{w}} = \underset{\boldsymbol{w}}{\operatorname{argmax}} \log p(\boldsymbol{w}|X, \Omega, k)$$
$$= \underset{\boldsymbol{w}}{\operatorname{argmax}} \log h(X|\boldsymbol{w}) - \frac{1}{2} \boldsymbol{w}^\top (I + \Lambda^{-1}) \boldsymbol{w}. \tag{8}$$

Crucially, $\hat{\boldsymbol{w}}$ must satisfy the stationarity condition

$$\hat{\boldsymbol{w}} = (I + \Lambda^{-1})^{-1} \nabla_{\boldsymbol{w}} \log h(X|\boldsymbol{w})|_{\boldsymbol{w}=\hat{\boldsymbol{w}}}, \tag{9}$$

where

$$\nabla_{\boldsymbol{w}} \log h(X|\boldsymbol{w})|_{\boldsymbol{w}=\hat{\boldsymbol{w}}} = 2 \sum_{i=1}^m \frac{\Phi(\boldsymbol{x}_i)}{\Phi(\boldsymbol{x}_i)^\top \hat{\boldsymbol{w}}}.$$

The approximate predictive mean is therefore

$$\hat{f}(\boldsymbol{x}^*) := \mathbb{E}[f(\boldsymbol{x}^*)|X, \Omega, k]$$
$$= \Phi(\boldsymbol{x}^*)^\top \hat{\boldsymbol{w}}$$
$$= \sum_{i=1}^m \frac{2}{\Phi(\boldsymbol{x}_i)^\top \hat{\boldsymbol{w}}} \cdot \Phi(\boldsymbol{x}_i)^\top (I + \Lambda^{-1})^{-1} \Phi(\boldsymbol{x}^*)$$
$$:= \sum_{i=1}^m \alpha_i \tilde{k}(\boldsymbol{x}_i, \boldsymbol{x}^*). \tag{10}$$

This reveals the same $\tilde{k}$ as (5). From (10) we have

$$\hat{\alpha}_i = 2/\hat{f}(\boldsymbol{x}_i).$$

Putting (9) and (10) into (8), we obtain

$$\hat{\boldsymbol{\alpha}} = \underset{\boldsymbol{\alpha}}{\operatorname{argmax}} \sum_{i=1}^m 2 \log |K_{i,:}\boldsymbol{\alpha}| - \frac{1}{2} \boldsymbol{\alpha}^\top \tilde{K} \boldsymbol{\alpha},$$

where $K = (k(\boldsymbol{x}_i, \boldsymbol{x}_j))_{ij}$ and $\tilde{K} = (\tilde{k}(\boldsymbol{x}_i, \boldsymbol{x}_j))_{ij}$. This is equivalent to Flaxman et al. (2017) (and the analogous section 3), even though here we did not appeal to the representer theorem.

#### 4.1.4. Predictive Variance

We now compute the $Q$ in (7). The Hessian term giving the inverse covariance becomes

$$\begin{aligned} Q^{-1} &= -\frac{\partial^2}{\partial \boldsymbol{w} \partial \boldsymbol{w}^\top} \log p(\boldsymbol{w}, X, \Omega, k)\Big|_{\boldsymbol{w}=\hat{\boldsymbol{w}}} \\ &= I + \Lambda^{-1} + W \\ W &= -\frac{\partial^2}{\partial \boldsymbol{w} \partial \boldsymbol{w}^\top} \log h(X|\boldsymbol{w})\Big|_{\boldsymbol{w}=\hat{\boldsymbol{w}}} \\ &= 2 \sum_{i=1}^{m} \frac{\Phi(\boldsymbol{x}_i)\Phi(\boldsymbol{x}_i)^\top}{(\Phi(\boldsymbol{x}_i)^\top \hat{\boldsymbol{w}})^2} \\ &:= VDV^\top, \end{aligned}$$

where $V_{:i} = \alpha_i \times \Phi(\boldsymbol{x}_i)$ and $D = \frac{1}{2} I \in \mathbb{R}^{m \times m}$. The approximate predictive variance can now be rewritten as an $m$-dimensional matrix expression using the identity $(Z + VDV^\top) = Z^{-1} - Z^{-1}V(V^\top Z^{-1} V + D^{-1})V^\top Z^{-1}$ with and $Z = I + \Lambda^{-1}$ along with a little algebra, to derive[2]

$$\begin{aligned} &\sigma^2(\boldsymbol{x}^*) := \mathrm{Var}\left[f(\boldsymbol{x}^*)|X, \Omega, k\right] \\ &= \Phi(\boldsymbol{x}^*)^\top Q\ \Phi(\boldsymbol{x}^*) \\ &= \tilde{k}(\boldsymbol{x}^*, \boldsymbol{x}^*) - \big(\tilde{k}(\boldsymbol{x}^*, X) \odot \boldsymbol{\alpha}\big) S^{-1} \big(\boldsymbol{\alpha}^\top \odot \tilde{k}(X, \boldsymbol{x}^*)\big), \end{aligned}$$

where $\odot$ is the Hadamard product, or element-wise multiplication, and $S := \big(\tilde{k}(X, X) \odot (\boldsymbol{\alpha}\boldsymbol{\alpha}^\top) + 2I\big)$.

#### 4.1.5. Predictive Distribution

Given the approximate predictive distribution $f(\boldsymbol{x}^*)|X, \Omega, k \sim \mathcal{N}(\hat{\boldsymbol{f}}(\boldsymbol{x}^*), \sigma^2(\boldsymbol{x}^*)) := \mathcal{N}(\mu, \sigma^2)$ and the relation $\lambda(\cdot) = \frac{1}{2} f^2(\cdot)$ it is straightforward to derive the corresponding[3] $\lambda(\boldsymbol{x}^*)|X, \Omega, k \sim \mathrm{Gamma}(\alpha, \beta)$ where the shape $\alpha = \frac{(\mu^2+\sigma^2)^2}{2\sigma^2(2\mu^2+\sigma^2)}$ and the scale $\beta = \frac{2\mu^2\sigma^2+\sigma^4}{\mu^2+\sigma^2}$.

#### 4.1.6. Marginal Likelihood

Letting $q(\hat{\boldsymbol{w}}, X|\Omega, k)$ be the Taylor expansion of $\log p(\boldsymbol{w}, X|\Omega, k)$ about the mode $\boldsymbol{w} = \hat{\boldsymbol{w}}$ and evaluating at $\hat{\boldsymbol{w}}$ gives, as linear and quadratic terms vanish,

$$\begin{aligned} &\log q(\hat{\boldsymbol{w}}, X|\Omega, k) = \log p(\hat{\boldsymbol{w}}, X|\Omega, k) \\ &= \log h(X|\hat{\boldsymbol{w}}) - \frac{1}{2} \hat{\boldsymbol{w}}^\top (I + \Lambda^{-1}) \hat{\boldsymbol{w}} - \frac{1}{2} \log |\Lambda| - \frac{N}{2} \log 2\pi. \end{aligned}$$

[2] Where *e.g.* $\tilde{k}(X, \boldsymbol{x}^*)$ is an $m \times 1$ matrix of evaluations of $\tilde{k}$.

[3] $\mathrm{Gamma}(x|\alpha, \beta)$ has p.d.f. $\frac{1}{\Gamma(k)\beta^k} x^{\alpha-1} \exp(-x/\beta)$.

Similarly to (18) we get approximate marginal likelihood

$$\begin{aligned} &\log p(X|\Omega, k) \approx \log q(\hat{\boldsymbol{w}}, X|\Omega, k) - \log q(\hat{\boldsymbol{w}}|X, \Omega, k) \\ &= \underbrace{\log h(X|\hat{\boldsymbol{w}})}_{-\sum_{i=1}^m \log 2\alpha_i^2} - \frac{1}{2}\Big(\underbrace{\hat{\boldsymbol{w}}^\top (I + \Lambda^{-1}) \hat{\boldsymbol{w}}}_{\boldsymbol{\alpha}^\top \tilde{k}(X,X) \boldsymbol{\alpha}} - \log |\Lambda| + \log |Q|\Big) \end{aligned} \tag{11}$$

We now use the determinant identity $|Z + VDV^\top| = |Z||D||V^\top Z^{-1} V + D^{-1}|$ with the same $Z$, $V$ and $D$ as subsubsection 4.1.4 to derive

$$\begin{aligned} &-\log |\Lambda| + \log |Q| = -\log |\Lambda| - \log |Z + VDV^\top| \\ &= -\log \big|\Lambda(I + \Lambda^{-1})\big| - \log \big|D^{-1} + V^\top Z^{-1} V\big| + c \\ &= \underbrace{\sum_{i=1}^{N} \log \frac{1}{1 + \lambda_i}}_{:= \mathcal{V}(k, \Omega, \mu)} - \log \Big|\tilde{k}(X, X) \odot (\boldsymbol{\alpha}\boldsymbol{\alpha}^\top) + 2I\Big| + c, \end{aligned} \tag{12}$$

where $c = m \log(2)$. $\mathcal{V}(k, \Omega, \mu)$ is the crucial ingredient, not accounted for by naïvely putting $\tilde{k}$ into subsection A.3.

## 5. Covariance Functions

To apply our inference scheme we need to compute:

1. The function $\tilde{k}$ from equation (10), studied recently by Flaxman et al. (2017) and earlier by Sollich & Williams (2005) as the *equivalent kernel.*

2. The associated term $\mathcal{V}(k, \Omega, \mu)$ from equation (12), required for the marginal likelihood.

This is often challenging for compact domains such as the unit hyper-cube. Such domains are crucial however, if we are to avoid the well-known edge-effects which arise from neglecting the fact that our data are sampled from, say, a two dimensional rectangle. In subsection 5.1 we provide a simple constructive approach to the case $\Omega = [0, 1]^d$. The following subsection 5.2 presents the general approximation scheme due to Flaxman et al. (2017), for the case when we have $k$ but not its Mercer expansion.

### 5.1. Thin-Plate Semi-norms on the Hyper-Cube

Consider the input domain $\Omega = [0, \pi]^d$ with Lebesgue measure $\mu$. A classical function regularisation term is the so called $m$-th order thin-plate spline semi-norm,

$$\begin{aligned} \langle f, g \rangle_{\mathcal{TP}(m)} &:= \sum_{|\alpha|=m} \frac{m!}{\prod_j \alpha_j!} \int_{\boldsymbol{x} \in \Omega} \frac{\partial^m f}{\partial \boldsymbol{x}^\alpha} \frac{\partial^m g}{\partial \boldsymbol{x}^\alpha} \mathrm{d}\mu(\boldsymbol{x}) \\ &= \langle f, \Delta^m g \rangle_{L^2(\Omega)} + \mathcal{B}. \end{aligned} \tag{13}$$

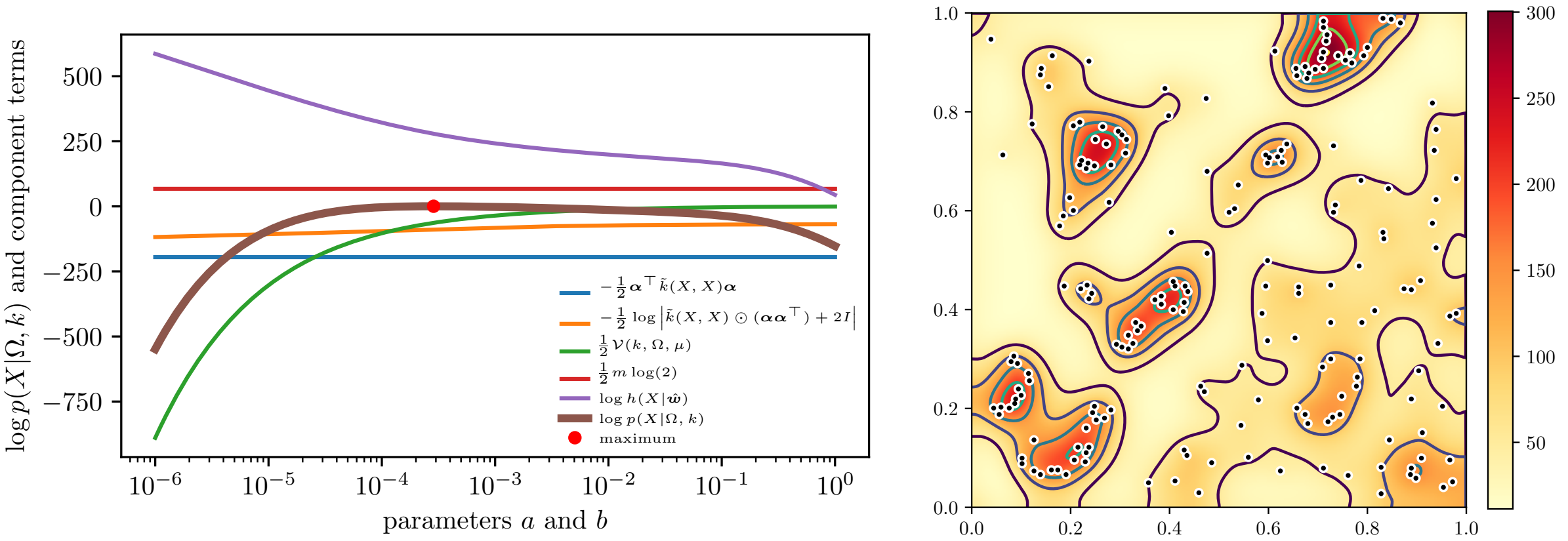

(a) Decomposition of the marginal likelihood

(b) Predictive mean intensity.

*Figure 2.* Modeling the Redwood dataset: (a) the log marginal likelihood $\log p(X|\Omega, k)$ along with its component terms from (11) and (12), as a function of the hyper-parameter $a$ from subsection 5.1; (b) the mean intensity corresponding to the maximum marginal likelihood parameters, with isolines at 50, 100, 150, 200 and 250. See subsection 6.5 for the details.

Here $\alpha$ is a multi-index running over all indices of total order $|\alpha| := \sum_j \alpha_j = m$, and the boundary conditions $\mathcal{B}$ come from formal integration (see *e.g.* Wahba (1990, section 2.4). We neglect $\mathcal{B}$ (for reasons explained shortly) and include the zero-th derivative to define

$$\langle f, g\rangle_{\mathcal{H}(k)} := \langle f, (a\Delta^m + b)g\rangle_{L^2(\Omega)}.$$

We may select the free parameters $a > 0$, $b > 0$ and $m \in \mathbb{Z}^+$ using the maximum marginal likelihood criterion. In general, it is challenging to obtain the expressions we require in closed form for arbitrary $d$, $\Omega$ and $m$. The analytical limit in the literature appears to be the case $m = 2$ with dimension $d = 1$ along with so-called Neumann boundary conditions (which impose a vanishing gradient on the boundary (Sommerfeld & Straus, 1949)). That $\tilde{k}$ has been derived in closed form as the reproducing kernel of an associated Sobolev space by Thomas-Agnan (1996).

We now present a simple but powerful scheme which sidesteps these challenges via a well chosen series expansion. Consider the basis function

$$\phi_\beta(\boldsymbol{x}) := (2/\pi)^{d/2}\prod_{j=1}^{d}\sqrt{1/2}^{[\beta_j=0]}\cos(\beta_j x_j),$$

where $\beta$ is a multi-index with non-negative (integral) values, and $[\cdot]$ denotes the indicator function (which is one if the condition is satisfied and zero otherwise). The $\phi_\beta$ form a convenient basis for our purposes. They are orthonormal:

$$\langle \phi_\beta, \phi_\gamma\rangle_{L^2(\Omega)} = [\beta = \gamma],$$

and also eigenfunctions of our regularisation operator with

$$(a\Delta^m + b)\phi_\beta = \Big(a\big(\sum_{j=1}^{d}\beta_j^2\big)^m + b\Big)\phi_\beta. \qquad (14)$$

Now if we restrict the function space to

$$\mathcal{H}(k) := \Big\{f = \sum_{\beta\geq\mathbf{0}} c_\beta\phi_\beta : \|f\|^2_{\mathcal{H}(k)} = \sum_{\beta\geq\mathbf{0}} c_\beta^2/\lambda_\beta < \infty\Big\},$$

then it is easily verified that the boundary conditions $\mathcal{B}$ in (13) vanish. This is a common approach to solving partial differential equations with Neumann boundary conditions (see *e.g.* Sommerfeld & Straus (1949)). By restricting in this way, we merely impose zero partial derivatives at the boundary, while otherwise enjoying the usual Fourier series approximation properties. Hence we can combine the reproducing property (4) with (13) and (14) to derive

$$k(\boldsymbol{x}, \boldsymbol{y}) = \sum_{\beta\geq\mathbf{0}} \lambda_\beta\phi_\beta(\boldsymbol{x})\phi_\beta(\boldsymbol{y}), \qquad (15)$$

where $\lambda_\beta := 1/(a\big(\sum_{j=1}^d \beta_j^2\big)^m + b)$.

The above covariance function is not required for our inference algorithm. Rather, the point is that since the basis is also orthonormal, we may substitute $\lambda_\beta$ and $\phi_\beta$ into (10) and (12) to obtain $\tilde{k}$ and $\mathcal{V}(k)$, as required.

**Series truncation.** We have discovered closed form expressions for $\tilde{k}$ only for $m \leq 2$ and $d = 1$. In practice we may truncate the series at any order and still obtain a valid model due to the equivalence with the linear model (6). Hence, a large approximation error (in terms of $\tilde{k}$) due to truncation may be irrelevant from a machine learning perspective, merely implying a different GP prior over functions. Indeed, the maximum marginal likelihood criterion based on subsubsection 4.1.6 may guide the selection of an appropriate truncation order, although some care needs to be taken in this case.

### 5.2. Arbitrary Covariances and Domains

Flaxman et al. (2017) suggested the following approximation for $\tilde{k}$, for the case when $k$ is known but the associated Mercer expansion is not. The approximation is remarkably general and elegant, and may even be applied to non-vectorial data by employing, say, a kernel function defined on strings (Lodhi et al., 2002). The idea is to note that the $\phi_i, \lambda_i$ pairs are eigenfunctions of the integral operator (see Rasmussen & Williams (2006) section 4.3)

$$T_k : \mathcal{H}(k) \to \mathcal{H}(k)$$
$$f \mapsto T_k f := \int_{\boldsymbol{x}\in\Omega} k(\boldsymbol{x}, \cdot) f(\boldsymbol{x}) p(\boldsymbol{x}) \,\mathrm{d}\boldsymbol{x},$$

where $p$ is related to $\mu$ of the previous subsection by $\mu(\boldsymbol{x}) = p(\boldsymbol{x})\,\mathrm{d}\boldsymbol{x}$. The Nyström approximation (Nyström, 1928) to $T_k$ draws $m$ samples $X$ from $p$ and defines $T_k^{(X)} g := \frac{1}{m} \sum_{\boldsymbol{x}\in X} k(\boldsymbol{x}, \cdot) g(\boldsymbol{x})$. Then the eigenfunctions and eigenvectors of $T_k$ may be approximated via the eigenvectors $\boldsymbol{e}_i^{(\mathrm{mat})}$ and eigenvalues $\lambda_i^{(\mathrm{mat})}$ of[2] $k(X, X)$, as

$$\phi_i^{(X)} := \sqrt{m}/\lambda_i^{(\mathrm{mat})} k(\cdot, X)\boldsymbol{e}_i^{(\mathrm{mat})}$$
$$\lambda_i^{(X)} := \lambda_i^{(\mathrm{mat})}/m.$$

These approximations may be used for $\tilde{k}$, as in (Flaxman et al., 2017), as well as our $\mathcal{V}(k, \Omega, \mu)$.

## 6. Experiments

### 6.1. Setup

**Evaluation** We use two metrics. The $\boldsymbol{\ell_2}$ **Error** is the squared difference to the ground truth w.r.t. the Lebesgue measure: $\int_{\boldsymbol{x}\in\Omega} (\lambda(\boldsymbol{x}) - \lambda_{\text{true}}(\boldsymbol{x}))^2 \,\mathrm{d}\boldsymbol{x}$. The **test log likelihood** is the logarithm of (1) at an independent test sample (one sample being a set of points, *i.e.* a sample from the process), which we summarise by averaging over a finite number of test sets (for real data where the ground truth intensity is unknown) and otherwise (if we have the ground truth) by the analytical expression $\mathbb{E}_{X\sim\mathrm{PP}(\lambda)}\left[\log p_{X\sim\mathrm{PP}(\hat{\lambda})}(X)\right] =$

$$\int_{\boldsymbol{x}\in\Omega} \left(\lambda(\boldsymbol{x}) \log \hat{\lambda}(\boldsymbol{x}) - \hat{\lambda}(\boldsymbol{x})\right) \mathrm{d}\boldsymbol{x},$$

where $PP(\lambda)$ is the process with intensity $\lambda$ (see the supplementary subsection A.1). This evaluation metric is novel in this context, yet more accurate and computationally cheaper than the sampling of *e.g.* (Adams et al., 2009).

**Decision Theory** The above metrics are functions of a single estimated intensity. In all cases we use the predictive mean intensity for evaluation. We demonstrate in subsection A.2 of the supplementary material that this is optimal

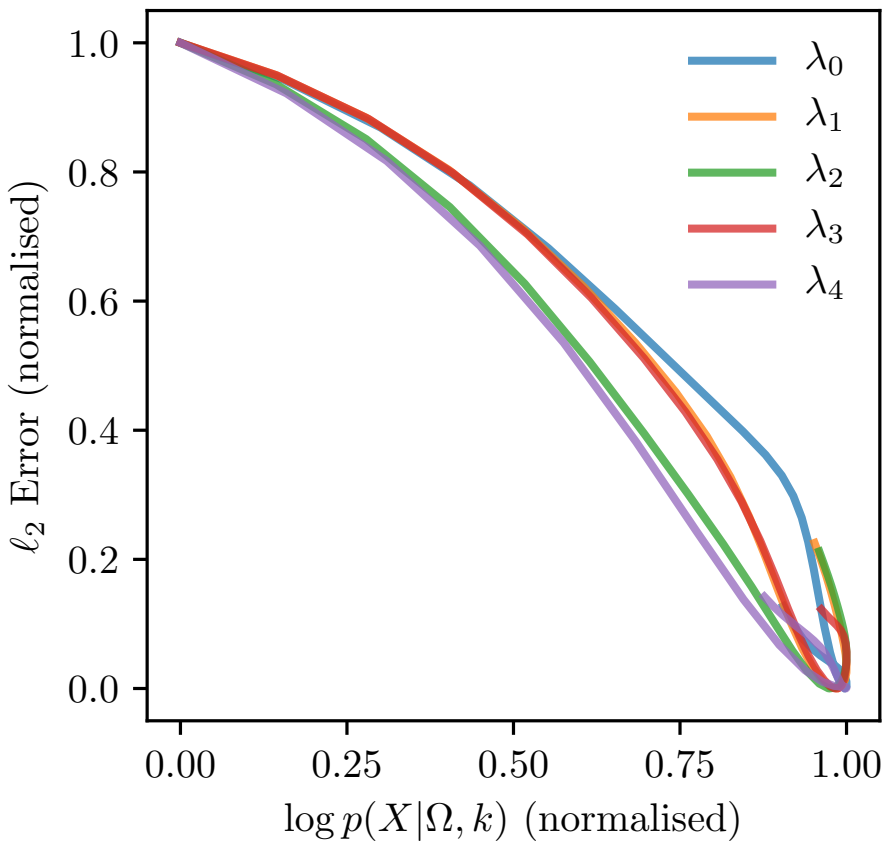

(a) $\ell_2$ error *vs.* marginal likelihood.

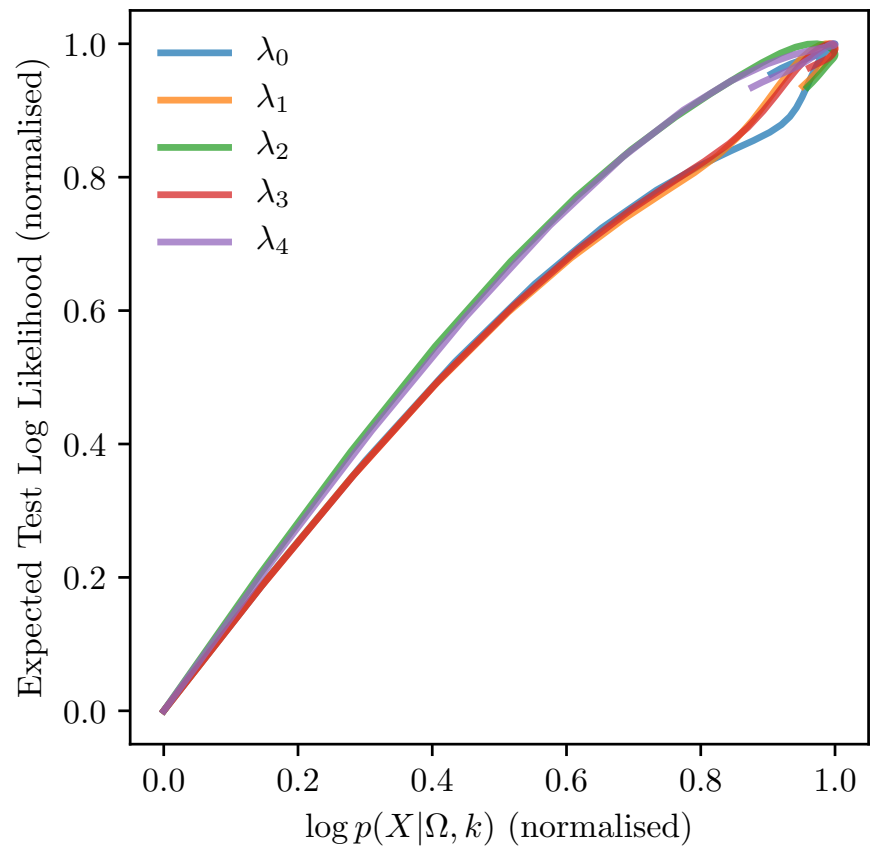

(b) Expected log-loss *vs.* marginal likelihood.

*Figure 3.* The relationship between the log marginal likelihood and the $\ell_2$ error (both scaled and shifted to $[0, 1]$), on the benchmark problems of subsection 6.2 — see figure 3 for the details.

for the expected test log likelihood evaluation (the $\ell_2$ error cases is similar as is trivial to show).

**Algorithms** We compare our new *Laplace Bayesian Point Process* (LBPP) with two covariances: the cosine kernel of subsection 5.1 with fixed $m = 2$ and hyper-parameters $a$ and $b$ (LBPP-Cos), and the Gaussian kernel $k(\boldsymbol{x}, \boldsymbol{z}) = \gamma^2 \exp(|\boldsymbol{x} - \boldsymbol{y}|^2 / (2\beta^2))$ with the method of subsection 5.2 (LBPP-G). We compared with the *Variational Bayesian Point Process* (VBPP) (Lloyd et al., 2015) using the same Gaussian kernel. LBPP-G and VBPP use a regular grid for $X$ (of subsection 5.2) and the inducing points, respectively. To compare timing we vary the number of *basis functions*, *i.e.* the number of grid points for LBPP-G and VBPP, and cosine terms for LBPP-Cos. We include the baseline *kernel smoothing with edge correction* (KS+EC) method (Diggle, 1985; Lloyd et al., 2015). All inference is performed with maximum marginal likelihood, except for KS+EC where we maximise the leave one out metric described in (Lloyd et al., 2015).

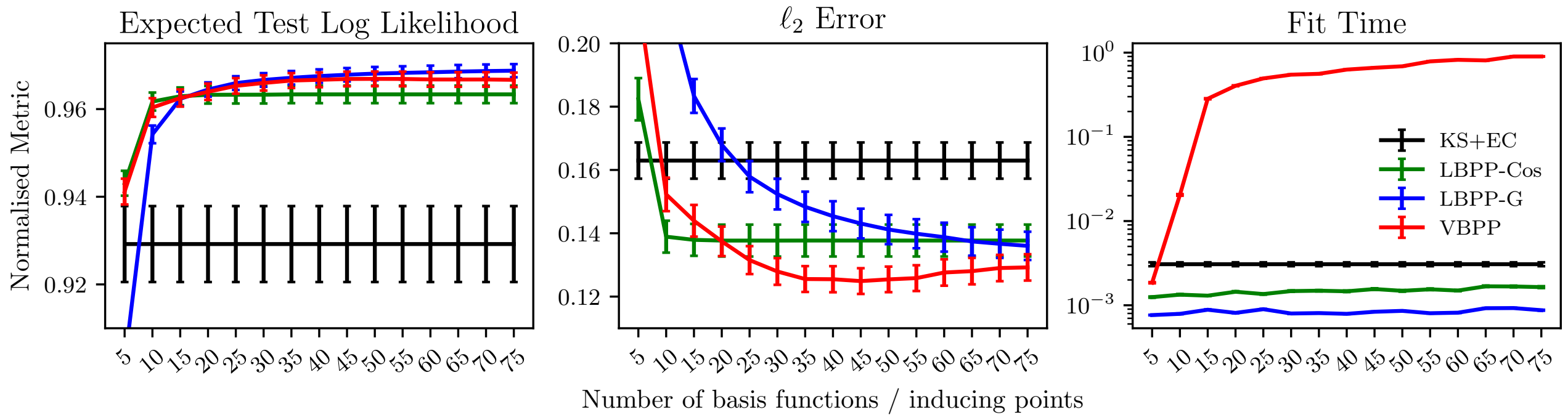

*Figure 4.* Mean ± one standard error performance on the toy problems of subsection 6.2 as a function of the number of basis functions (with KS+EC results replicated along the horizontal axis for comparison). The vertical axes of each figure are normalised scores: see subsubsection 6.2.2 for the details.

### 6.2. 1D Toy Examples

We drew five toy intensities, $\lambda_0, \lambda_2, \ldots, \lambda_4$ as $\frac{1}{2}f^2$ where $f$ was sampled from the GP of Gaussian covariance (defined above) with $\gamma = 5$ and $\beta = 0.5$. Figure 1 depicts $\lambda_0$ — see the caption for a description. The remaining test functions are shown in figure 6 of the supplementary material.

#### 6.2.1. Model Selection

As the marginal likelihood $\log p(X|\Omega, k)$ is a key advantage of our method over the non-probabilistic approach of Flaxman et al. (2017), we investigated its efficacy for model selection. Figure 3 plots $\log p(X|\Omega, k)$ against our two error metrics, both rescaled to $[0, 1]$ for effective visualisation, based on a single training sample per test function. We observe a strong relationship, with larger values of $\log p(X|\Omega, k)$ generally corresponding to lower error. This demonstrates the practical utility of both the marginal likelihood itself, and our Laplace approximation to it.

#### 6.2.2. Evaluation

We sampled 100 training sets from each of our five toy functions. Figure 4 shows our evaluation metrics along with the fitting time as a function of the number of basis functions. For visualisation all metrics (including fit time) are scaled to $[0, 1]$ by dividing by the maximum for the given test function, over data replicates and algorithms. LBBP-G and and VBPP achieve the best performance, but our LBPP-G is two orders of magnitude faster. Our KS+EC implementation follows the methodology of Lloyd et al. (2015): we fit the kernel density bandwidth using average *leave one out* log likelihood. This involves a quadratic number of log p.d.f. of the truncated normal calculations, and log-sum-exp calculations, both of which involve large time constants, but are asymptotically superior to the other methods we considered. LBBP-Cos is slightly inferior in terms of expected test log likelihood, which is expected due to the toy functions having been sampled according to the same Gaussian kernel of LBPP-G and VBPP (as well as the density estimator of KS+EC).

### 6.3. Real Data

We compared the methods on three real world datasets,

- *coal:* 190 points in one temporal dimension, indicating the time of fatal coal mining accidents in the United Kingdom, from 1851 to 1962 (Collins, 2013);
- *redwood:* 195 California redwood tree locations from a square sampling region (Ripley, 1977);
- *cav:* 138 caveolae locations from a square sampling region of muscle fiber (Davison & Hinkley, 2013).

### 6.4. Computational Speed

Similarly to subsubsection 6.2.2 we evaluate the fitting speed and statistical performance *vs.* number of basis functions — see figure 5. We omit the $\ell_2$ error as the ground truth is unknown. Instead we generate 100 test problems by each time randomly assigning each original datum to either the training or the testing set with equal probability. Again we observe similar predictive performance of LBPP and VBPP, but with much faster fit times for our LBPP. Interestingly LBPP-Cos slightly outperform LBPP-G.

### 6.5. 2D California Redwood Dataset

We conclude by further investigating the *redwood* dataset. Once again we employed the ML-II procedure to determine $a$ and $b$, fixing $m = 2$, for the covariance function of subsection 5.1, using the lowest 32 cosine frequencies in each dimension for a total of $N = 32^2$ basis functions in the expansion (15). For ease of visualisation we also fixed $a = b$.

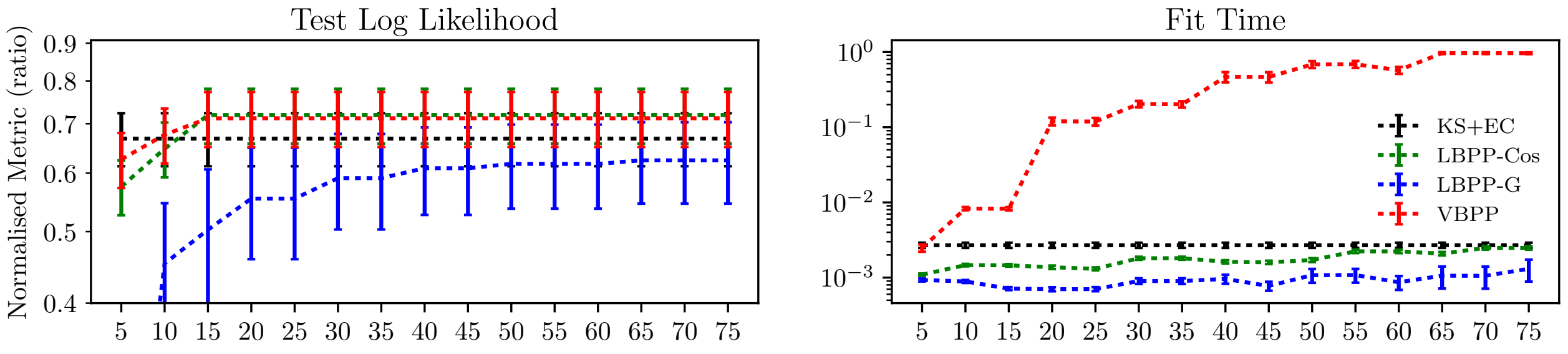

(a) *cas* dataset

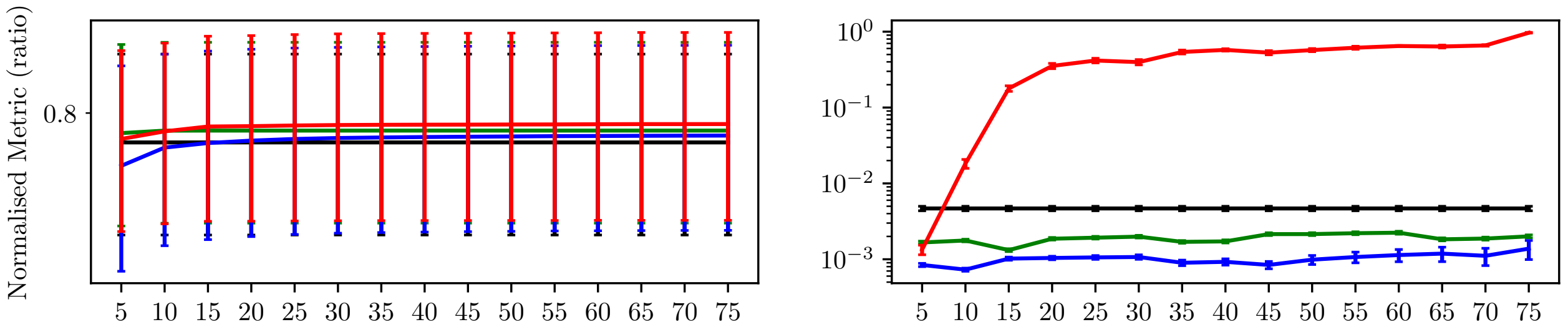

(b) *coal* dataset

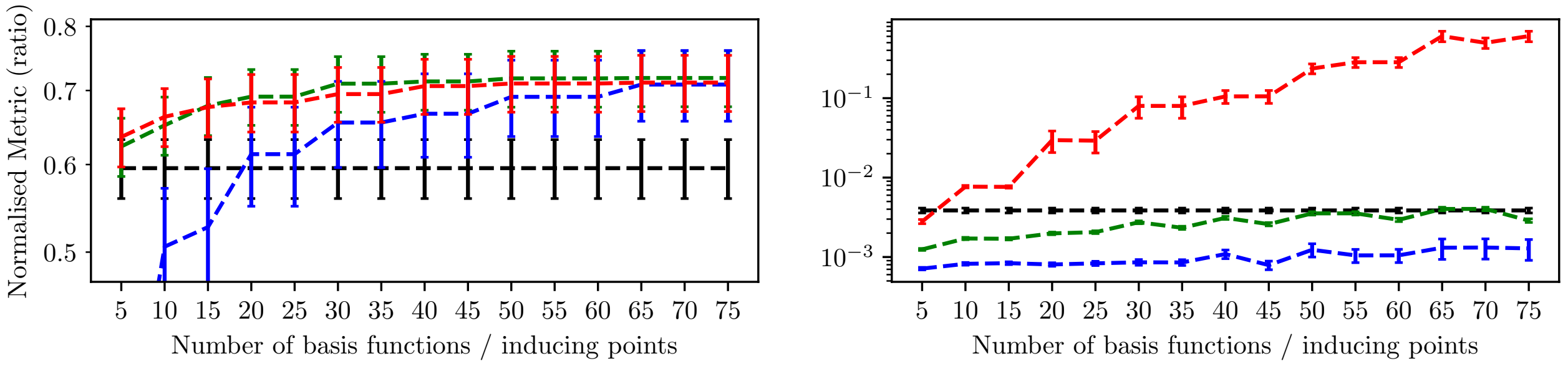

(c) *redwood* dataset

*Figure 5.* Mean ± one standard error performance of the different methods on real data, as a function of the number of basis functions (with KS+EC results replicated along the horizontal axis for comparison). See subsection 6.4 for the details.

Figure 2 plots the results, including a decomposition of the log marginal likelihood $\log p(X|\Omega, k)$ and a visualisation of the predictive mean. The mean function strongly resembles the result presented by Adams et al. (2009), where computationally expensive MCMC was employed.

The decomposition of the marginal likelihood on the left of figure 2 provides insight into the role of the individual terms in (11) and (12) which make up $\log p(X|\Omega, k)$. In particular, the term $\mathcal{V}(k, \Omega, \mu)$ from (12) acts as a regulariser, guarding against over-fitting, and balancing against the data term $h$ of (11).

## 7. Conclusion

We have discussed the permanental process, which places a Gaussian Process prior over the square root of the intensity function of the Poisson process, and derived the equations required for *empirical Bayes* under a Laplace posterior approximation. Our analysis provides 1) an alternative derivation and probabilistic generalization of (Flaxman et al., 2017), and 2) an efficient and easier to implement alternative which does not rely on inducing inputs (but rather reproducing kernel Hilbert space theory), to the related Bayesian approach of Lloyd et al. (2015). This further demonstrates, in a new way, the mathematical convenience and practical utility of the permanental process formulation (in comparison with say log Gaussian Cox processes).

## Acknowledgements

Thanks to Young Lee, Kar Wai Lim and Cheng Soon Ong for useful discussions. Adrian is supported by the Australian Research Council (ARC) via a Discovery Early Career Researcher Award (DE-120102873).

# A. Supplementary Material

Accompanying the submission *Fast Bayesian Intensity Estimation for the Permanental Process*.

## A.1. Exact Expected Log Loss

We evaluate our estimated $\hat{\lambda}$ using the expectation under the true $\mathrm{PP}(\lambda)$ of the log likelihood under $\mathrm{PP}(\hat{\lambda})$, where PP is the Poisson process. Adams et al. (2009) approximate this quantity using Monte Carlo, employing numerical integration for (1). It turns out that for the computational cost of one such numerical integration, we may compute the expected loss using standard results for Lévy processes (Cont & Tankov, 2004). An elementary self contained argument runs as follows:

$$\begin{aligned}
\mathbb{E}_{X\sim\mathrm{PP}(\lambda)}\left[\log p_{X\sim\mathrm{PP}(\hat{\lambda})}(X)\right] &= \mathbb{E}_{\mathrm{card}(X)}\left[\mathbb{E}_{X\sim\mathrm{PP}(\lambda)|\,\mathrm{card}(X)}\left[\log p_{X\sim\mathrm{PP}(\hat{\lambda})}(X)\right]\right] \\
&= \mathbb{E}_{\mathrm{card}(X)}\left[\mathrm{card}(X)\left(\log\hat{\Lambda}(\Omega) + H(\lambda,\hat{\lambda})\right) - \hat{\Lambda}(\Omega)\right] \\
&= \Gamma(\Omega)\left(\log\hat{\Lambda}(\Omega) + H(\lambda,\hat{\lambda})\right) - \hat{\Lambda}(\Omega) \\
&= \int_{\boldsymbol{x}\in\Omega}\left(\lambda(\boldsymbol{x})\log\hat{\lambda}(\boldsymbol{x}) - \hat{\lambda}(\boldsymbol{x})\right)\mathrm{d}\boldsymbol{x},
\end{aligned}$$

where $\Omega$ is the sampling domain, $H(\lambda,\hat{\lambda}) := \int_{\boldsymbol{x}\in\Omega}\frac{\lambda(\boldsymbol{x})}{\Lambda(\Omega)}\log\frac{\hat{\lambda}(\boldsymbol{x})}{\hat{\Lambda}(\Omega)}\,\mathrm{d}\boldsymbol{x}$ is the cross-entropy between the probability density functions proportional to $\lambda$ and $\hat{\lambda}$ and we recall $\Lambda(S) := \int_{x\in S}\lambda(x)\,\mathrm{d}x$. The first line is the tower law of expectation. To see the second line, note that we may sample $X\sim\mathrm{PP}(\lambda)$ by first sampling $\mathrm{card}(X)\sim\mathrm{Poisson}(\Lambda(\Omega))$, and then drawing each element of $X$ according to the probability density proportional to $\lambda$. The third line uses the Poisson expectation $\mathbb{E}_{\mathrm{card}(X)}\left[\mathrm{card}(X)\right] = \Gamma(\Omega)$ and the fourth some simple algebra.

As an aside, we may therefore write the Kullback-Leibler divergence in a form resembling that for probability distributions:

$$\begin{aligned}
D_{\mathrm{KL}}\left(\mathrm{PP}(f)\|\,\mathrm{PP}(g)\right) &= \mathbb{E}_{X\sim\mathrm{PP}(\lambda)}\left[\log p_{X\sim\mathrm{PP}(\lambda)}(X) - \log p_{X\sim\mathrm{PP}(\hat{\lambda})}(X)\right] \\
&= \int_{\boldsymbol{x}\in\Omega}\left(f(\boldsymbol{x})\log\frac{f(\boldsymbol{x})}{g(\boldsymbol{x})} + g(\boldsymbol{x}) - f(\boldsymbol{x})\right)\mathrm{d}\boldsymbol{x}.
\end{aligned}$$

## A.2. Bayesian Decision Theory for the Expected Log Loss

To determine the intensity function which maximises the expected log likelihood we define the loss

$$\ell(\lambda,\lambda') := \mathbb{E}_{n_i\sim N(B_i),i=1,2,\ldots,m|\lambda}\log p\left(N(B_i)=n_i, i=1,2,\ldots,m|\lambda'\right)$$

where $N(B_i)$ is the random variable representing the number of points in the set $B_i\subseteq\Omega$, $\Omega$ is the domain of the process and we recall $\Lambda(S) := \int_{x\in S}\lambda(x)\,\mathrm{d}x$. It is well known that (Baddeley, 2007)

$$p(N(B_i)=n_i, i=1,2,\ldots,m|\lambda) = \prod_i\frac{\Lambda(B_i)^{n_i}}{n_i!}\exp(-\Lambda(\Omega)).$$

Bayesian decision theory considers the expected loss

$$L(\lambda') := \mathbb{E}_{\lambda|D}\left[\ell(\lambda,\lambda')\right],$$

where the expectation is with respect to the posterior predictive distribution given the data $D$. Combining these expressions and assuming without loss of generality that $\Omega = \bigcup_i B_i$ yields

$$L(\lambda') = \mathbb{E}_{\lambda|D}\left[\mathbb{E}_{n_i\sim N(B_i),i=1,2,\ldots,m|\lambda}\left[\sum_i\left(n_i\log\Lambda'(B_i) - \log(n_i!) - \Lambda'(B_i)\right)\right]\right].$$

The optimal choice is $\Lambda^* := \mathrm{argmax}_{\lambda'}\,L(\lambda')$, so by stationarity

$$\begin{aligned}
\lambda^*(B_i) &= \mathbb{E}_{\lambda|D}\left[\mathbb{E}_{n_i\sim N(B_i)|\lambda}\left[n_i\right]\right] \\
&= \mathbb{E}_{\lambda|D}\left[\Lambda(B_i)\right],
\end{aligned}$$

and so $\lambda^* = \mathbb{E}_{\lambda|D}[\lambda]$, the expectation of the posterior predictive distribution.

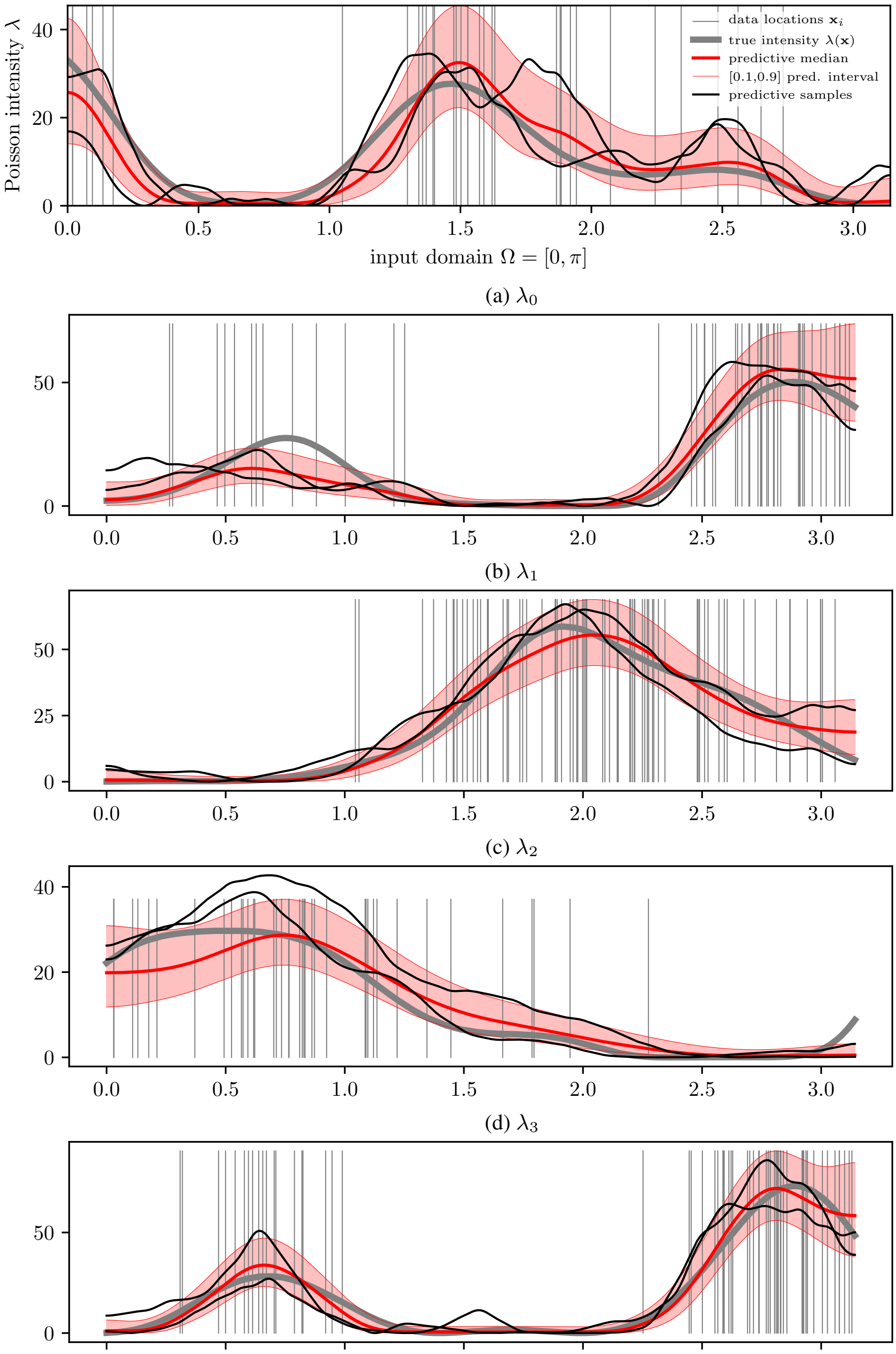

(a) $\lambda_0$

(b) $\lambda_1$

(c) $\lambda_2$

(d) $\lambda_3$

(e) $\lambda_4$

*Figure 6.* Predictive distributions for the test problems of subsection 6.2.

### A.3. Standard Laplace Approximations for the GP

Following *e.g.* (Rasmussen & Williams, 2006), assume that we are given an independent and identically distributed sample $\{(\boldsymbol{x}_i, y_i)\}_{1 \le i \le m}$, and the goal is to estimate $p(y|\boldsymbol{x})$. Let the true joint in $\boldsymbol{f} = (f(\boldsymbol{x}_i))_i, \boldsymbol{y} = (y(\boldsymbol{x}_i))_i$ be

$$\begin{aligned}\log p(\boldsymbol{y}, \boldsymbol{f}|X, k) &= \log p(\boldsymbol{y}|\boldsymbol{f}) + \log p(\boldsymbol{f}|X, k) \\ &= \log p(\boldsymbol{y}|\boldsymbol{f}) - \frac{1}{2}\boldsymbol{f}^\top K^{-1}\boldsymbol{f} - \frac{1}{2}\log|K| - \frac{m}{2}\log 2\pi,\end{aligned}$$

where $K = (k(\boldsymbol{x}_i), \boldsymbol{x}_j)_{ij}$ and $X = (\boldsymbol{x}_1, \boldsymbol{x}_2, \ldots, \boldsymbol{x}_m)$. The Laplace approximation fits a normal to the posterior,

$$\begin{aligned}\log p(\boldsymbol{f}|\boldsymbol{y}, X) &\approx \log \mathcal{N}(\boldsymbol{f}|\hat{\boldsymbol{f}}, Q) \\ &= -\frac{1}{2}(\boldsymbol{f} - \hat{\boldsymbol{f}})^\top Q^{-1}(\boldsymbol{f} - \hat{\boldsymbol{f}}) - \frac{1}{2}\log|Q| - \frac{m}{2}\log 2\pi \\ &:= \log q(\boldsymbol{f}|\boldsymbol{y}, X).\end{aligned}$$

$\hat{\boldsymbol{f}}$ and $Q$ come from a second order approximation of the log posterior at its mode, *i.e.*

$$\begin{aligned}\hat{\boldsymbol{f}} &= \operatorname*{argmax}_{\boldsymbol{f}} p(\boldsymbol{y}|\boldsymbol{f}, X) \\ &= \operatorname*{argmax}_{\boldsymbol{f}} p(\boldsymbol{y}, \boldsymbol{f}|X) \\ Q^{-1} &= -\left.\frac{\partial^2}{\partial \boldsymbol{f} \partial \boldsymbol{f}^\top} \log p(\boldsymbol{y}, \boldsymbol{f}|X)\right|_{\boldsymbol{f}=\hat{\boldsymbol{f}}} \\ &= K^{-1} + W \\ W_{ii} &= -\left.\frac{\partial^2}{\partial f_i^2} \log p(y_i|f_i)\right|_{f_i=\hat{f}_i}\end{aligned}$$

Taylor expanding $\log p(\boldsymbol{y}, \boldsymbol{f}|X)$ at $\boldsymbol{f} = \hat{\boldsymbol{f}}$,

$$\begin{aligned}\log p(\boldsymbol{y}, \boldsymbol{f}|X) &\approx \log p(\boldsymbol{y}, \hat{\boldsymbol{f}}|X) - \frac{1}{2}(\boldsymbol{f} - \hat{\boldsymbol{f}})^\top Q^{-1}(\boldsymbol{f} - \hat{\boldsymbol{f}}) \qquad (16) \\ &= \log p(\boldsymbol{y}|\boldsymbol{f} = \hat{\boldsymbol{f}}) - \frac{1}{2}\hat{\boldsymbol{f}}^\top K^{-1}\hat{\boldsymbol{f}} - \frac{1}{2}\log|K| - \frac{m}{2}\log 2\pi - \frac{1}{2}(\boldsymbol{f} - \hat{\boldsymbol{f}})^\top Q^{-1}(\boldsymbol{f} - \hat{\boldsymbol{f}}) \\ &:= \log q(\boldsymbol{y}, \boldsymbol{f}|X)\end{aligned}$$

Now

$$\log \int \exp(-\frac{1}{2}\boldsymbol{x}^\top H^{-1}\boldsymbol{x}) d\boldsymbol{x} = \frac{m}{2}\log 2\pi + \frac{1}{2}\log|H|$$

So we get the approximate marginal likelihood

$$\begin{aligned}\log Z &:= \log p(\boldsymbol{y}|X) \\ &\approx \log \int q(\boldsymbol{y}, \boldsymbol{f}|X) d\boldsymbol{f} \\ &= \log p(\boldsymbol{y}|\boldsymbol{f} = \hat{\boldsymbol{f}}) - \frac{1}{2}\hat{\boldsymbol{f}}^\top K^{-1}\hat{\boldsymbol{f}} - \frac{1}{2}\log|K| - \frac{1}{2}\log|K^{-1} + W| \\ &= \log p(\boldsymbol{y}|\boldsymbol{f} = \hat{\boldsymbol{f}}) - \frac{1}{2}\hat{\boldsymbol{f}}^\top K^{-1}\hat{\boldsymbol{f}} - \frac{1}{2}\log|I + KW| \qquad (17)\end{aligned}$$

This is a standard textbook approach (Rasmussen & Williams, 2006), but we can get the same approximation via

$$\log p(\boldsymbol{y}|X) \approx \log q(\boldsymbol{y}, \hat{\boldsymbol{f}}|X) - \log q(\hat{\boldsymbol{f}}|\boldsymbol{y}, X), \qquad (18)$$

since the right hand side is true for all $\boldsymbol{f}$, not just $\hat{\boldsymbol{f}}$. Hence we need only subtract the approximate log likelihoods as above. By evaluating at $\hat{\boldsymbol{f}}$, the second r.h.s. term in (16), vanishes immediately.